\documentclass[twocolumn,showpacs,preprintnumbers,amsmath,amssymb]{revtex4}
\usepackage{graphicx}% Include figure files
\usepackage{dcolumn}% Align table columns on decimal point
\usepackage{bm}% bold math
\usepackage{color}
\begin{document}
\input epsf

\title{Cytoskeleton confinement of red blood cell membrane fluctuations}
\author{N. Gov, A.G. Zilman and S. Safran}
\address{Department of Materials and Interfaces,
The Weizmann Institute of Science,\\
P.O.B. 26, Rehovot, Israel 76100}

\begin{abstract}
We analyze both the static and dynamic fluctuation spectrum of the
red-blood cell in a unified manner, using a simple model of the
composite membrane. In this model, the two-dimensional spectrin
network that forms the cytoskeleton is treated as a rigid shell
which is located at some constant average separation from the
lipid bilayer. The cytoskeleton thereby confines both the static
and dynamic fluctuations of the lipid bilayer. The predictions of
the model account for the wavevector and frequency dependence of
the experimental data. The observed amplitude of the thermal
fluctuations is related to effects of ATP-driven fluctuations.
\end{abstract}


\maketitle

A long-standing problem in the study of red blood cell (RBC)
structure is the simultaneous softness of its membrane observed by
thermal fluctuations \cite{sackman95}, and the strong shear
elasticity found in static deformation experiments, such as
micropipette aspiration \cite{pipette} and electrodeformation
\cite{sackman84}. The membrane itself is a composite structure
\cite{sackman94} with an outer, gel-like extracellular network of
long sugar molecules (thought to be irrelevant to the structural
strength), a mixed lipid/protein bilayer and an attached,
intracellular network. Previous theoretical models of this
membrane treated it as a single, polymerized network with the
combined curvature bending modulus of the lipid bilayer $\kappa$
and the shear rigidity of the cytoskeleton $\mu$
\cite{peterson92}. Such models were successful in describing the
response of the membrane in static {\it deformation} experiments,
which give $\mu\sim 10^{-5}-10^{-6}$J/m$^2$
\cite{pipette,sackman84,discher01}. However, comparing these
models to the {\it fluctuation} data, leads to the conclusion that
the membrane behaves as if the shear modulus vanishes $\mu \sim 0$
\cite{brochard,peterson92}. This surprising conclusion comes from
the shape and amplitude analysis of the longest wavelength shape
fluctuations \cite{sackman92,sackman95}. Various ideas have been
raised in order to account for this observation, the main
suggestion being that ATP-driven structural rearrangement in the
spectrin network \cite{discher01,sackman94,sackman95} relaxes the
shear-like deformations.

The previous studies were concerned with the shape fluctuations of
largest wavelength. Here, we focus on the fluctuation spectrum, at
the length-scales of $1-0.01\mu$m, where the effect of the
cytoskeleton is observed \cite{sackman87}. The important question
is to what extent the cytoskeleton effects are distinguishable
from the fluctuations of a free, closed bilayer. We show that for
a consistent description of both the static and the dynamic
fluctuation spectrum we {\em must} include the confining effects
of the cytoskeleton. A simple model where the only significant
effect of the spectrin network is to confine the lipid bilayer
membrane, consistently describes both the spatial and temporal
spectra of the thermal fluctuations of the RBC membrane. We
discuss how active processes that effectively increase the
temperature of the membrane, can be accounted for within this
description.

The curvature bending modulus $\kappa$ of the lipid bilayer
\cite{seifert} is deduced from measurements of the amplitude of
thermal fluctuations at the smallest measured wavelengths
\cite{sackman87} $\kappa\simeq 2\pm1\times10^{-20}$J. The RBC
cytoskeleton is a two-dimensional, roughly triangular, network of
spectrin proteins \cite{bennett}, that is attached to the lipid
bilayer at the nodes and at additional, random sites along the
spectrin polymers. The cytoskeleton is well described as a network
of entropic springs, of length $l\sim80$nm
\cite{bennett,sackmangreenbook}, with an effective spring
constant: $\mu\simeq4\times10^{-6}$J/m$^2$, which is close to the
measured static shear modulus: $\sim6\times10^{-6}$J/m$^2$
\cite{pipette}. Compared with the lipid bilayer bending modulus
$\kappa$, the curvature bending modulus of the cytoskeleton is
negligible \cite{sackman95,landaumcdonald01discher98}:
$\kappa_{cyto}\sim \mu w^2\sim 10^{-21}-10^{-22}$J for a
cytoskeleton thickness of $w\sim100-500$\AA.

We now begin by analyzing the measured spatial correlations of the
fluctuations \cite{sackman87}, and describe the effects of the
cytoskeleton on the bilayer in terms of continuum mechanics. This
is feasible since the cytoskeleton forms a rather open mesh that
is attached to the bilayer at discrete points with a contact area
that is small ($\sim1$nm) compared to the inter-node distance
($\sim100$nm). In a coarse grained picture, we can describe the
thermal fluctuations of the bilayer using a continuum model of the
free energy functional
\begin{equation}
F\simeq \int dS\left[\frac{1}{2}\sigma\left(\nabla
h\right)^2+\frac{1}{2}\kappa\left(\nabla^2
h\right)^2+\frac{1}{2}\gamma h^2 \right] \label{freeenergymem1}
\end{equation}
that includes the bending and effective surface tension energy of
the bilayer (see for example \cite{safran}) in terms of the normal
displacement $h$. The surface tension coefficient $\sigma$ arises
from the constraint of constant surface area of the bilayer in the
closed geometry of the RBC \cite{sackman97,seifert}:
$\sigma_0=\alpha\kappa/R^2\sim 1\times10^{-9}$J/m$^2$ (taking
$R\sim 4\mu$m as the RBC radius, $\alpha=1$). The constant can
vary $\alpha\approx 1-100$.

The last term in (\ref{freeenergymem1}) is mathematically
equivalent to a Lagrange multiplier that constrains the mean
square amplitude of bilayer fluctuations to be equal to:
$d^2=k_{B}T/8(\gamma \kappa)^{1/2}$, for an infinite bilayer. This
term describes the effect of the cytoskeleton on the bilayer
through a harmonic potential that maintains an average separation
$d$ (of order $w$) between the lipid bilayer and the cytoskeleton
\cite{safran}, here treated as a separated, infinitely rigid
shell, that does not participate in the thermal fluctuations. The
discrete contacts that maintain the constant average separation
are not specifically described in this continuum model; in a
coarse-grained picture, these contacts are the physical origin of
the constraint (potential) that determines the average
membrane-spectrin network separation.

The attachment of the cytoskeleton to the bilayer also causes
stretching and undulations of the bilayer \cite{sackman90}, partly
due to steric repulsion between the spectrin and the bilayer
around the point of attachment \cite{lipowsky}. Balancing the
spectrin stretching with a local curvature of the bilayer, results
in a membrane with undulations of wavelength
\cite{sackman90,steric} $L\sim\sqrt{\kappa/\mu}\sim100-200$nm and
amplitude $\sim10$nm($\sim w$)
\cite{sackman95,landaumcdonald01discher98}. In our confining-shell
model this length-scale is related to the potential-induced
persistence length of the bilayer
$\xi_{0}=(\kappa/\gamma)^{1/4}\sim L$ \cite{safran}, i.e. the
wavelength below which the bilayer is freely fluctuating.

We now calculate the spatial correlations for a two-dimensional,
flat bilayer, since for all but the largest wavelengths $\lambda$,
the surface of the RBC is relatively flat: $50nm<\lambda<1\mu m <
R\sim 4\mu$m. We determine the values of $\sigma$ and $\gamma$ by
fitting to the experimental data. From Eq. (\ref{freeenergymem1})
the equal-time (static) correlations of the normal deflections of
the bilayer can be written \cite{safran,sackman92}
\begin{equation}
\langle h_{q}h_{-q}\rangle=\frac{k_{B}T}{\kappa_q q^{4}}, \  \
\kappa_q=\kappa+\sigma q^{-2} + \gamma q^{-4}
 \label{statcor}
\end{equation}

In the inset of Fig.1 we plot the measured value of $\kappa_q$
\cite{sackman87} in the form $(\kappa_q/\kappa-1)^{-1}$ as a
function of the normalized wavevector $(qd)^4$ (where $d$ is
determined by fitting the data to obtain $\gamma \propto d^{-4}$).
From the linear slope in the limit of $q\rightarrow 0$ we find the
values of the parameter $\gamma=7.5,1.0\times 10^{7}$J/m$^4$, for
the two cells measured. These values correspond to mean amplitudes
$d\simeq 200,350$\AA \ and
$\xi_{0}=(\kappa/\gamma)^{1/4}=130,220$nm respectively. At larger
values of $q$ there is a noticeable deviation from a straight
line, which arises from the effective surface tension
$\sigma\sim7.7,2.8\times10^{-7}$J/m$^2$ for the two cells. Note
that surface tension alone, without the confining effect of the
cytoskeleton (i.e. $\gamma=0$), does not fit the data (dash-dot
line, Fig.1 inset). In addition, there is a rather abrupt change
at the crossover wavevector $q_{0}=1/\xi_{0}$ (indicated by the
vertical dashed lines in Fig.1), above which the data are better
described using $\sigma\simeq\sigma_0 \sim
1.4\times10^{-9}$J/m$^2$ (solid lines in Fig.1).

The measured surface tension is consistent with:
$\sigma\simeq\kappa/\xi_{0}^2$ (of order $\sim\mu/10$). This
expression gives the effect of bilayer shape constraint due to the
static undulations of lateral size $\xi_{0}\times \xi_{0}$,
described above. Indeed at length-scales shorter then $\xi_{0}$,
there is no stretching of the cytoskeleton (Fig.1). Note that the
effective surface tension of a closed bilayer is a very sensitive
function of the excess area of the bilayer \cite{seifert,sens},
which is affected by the induced undulations. The spread in the
measured parameters may be due to natural variations in the
cytoskeleton network of normal RBC cells.

There is a qualitative difference in power-law of the wavevector
dependence of $\kappa_q$ for RBC and empty vesicles
\cite{sackman97}. The vesicles are well described (Fig.2) by
equations (\ref{freeenergymem1},\ref{statcor}) with $\gamma=0$,
and an effective surface tension: $\sigma_{vesicle}
\simeq\kappa/R^2\sim 2\times10^{-10}$J/m$^2$, where $R\sim 27\mu$m
and $\kappa\sim1.3\times 10^{-19}$J are the vesicle radius and
bending modulus respectively. Both the RBC's and vesicles data
collapse when the wavelength is scaled by the r.m.s amplitude $d$
(Fig.2). For the vesicles of diameter $\sim50\mu$m
\cite{sackman97} the r.m.s. thermal amplitudes are
$d\simeq1-1.5\mu$m (note that here $d$ is not related to
confinement). The good scaling of the data indicates that there is
indeed a single important length-scale in the problem, namely the
persistence length $\xi_0$, that determines all the parameters
appearing in the free energy ($\gamma$ and $\sigma$) and the r.m.s
amplitude $d$.

We now use the same simple model of spectrin confinement of the
bilayer to describe the temporal correlations of the membrane
fluctuations. The shape fluctuations of the RBC membrane are
driven by both thermal and metabolic energies. The active
fluctuations have a frequency spectrum that is confined to the
range 0.3-1Hz \cite{atprafi}. For higher frequencies, our analysis
shows that the active processes can be accounted for by an
increase in the effective temperature of the fluctuations
\cite{atprafi,prost}. The temporal height-height correlation
function \cite{brochard,anton} for a flat bilayer at a distance
$D_{\omega}$ from a rigid wall, is
\begin{equation}
\langle h_{q}(t)h_{-q}(0)\rangle=\frac{k_{B}T}{\kappa_q q^{4}}
e^{-\omega(q)t}
 \label{timecor}
\end{equation}
where $\kappa_q$ is given in (\ref{statcor}). The hydrodynamic
interaction (Oseen interaction kernel \cite{landauhyd}) has a
modified form: $\Lambda(q)=(1/4\eta q)\left[1-(1+2 q
D_{\omega})e^{-2 q D_{\omega}}\right]$ ($\Lambda_{f}(q)=1/4\eta q$
for a free bilayer), so that the relaxation frequency $\omega(q)$
is
\begin{equation}
\omega(q)=\left[\frac{1}{4 \eta}\left(\kappa q^3 +\sigma q+
\frac{\gamma}{q}\right)\right] \left[1-(1+2 q D_{\omega})e^{-2 q
D_{\omega}}\right] \label{omega}
\end{equation}
where $\eta\sim 3\eta_{water}$ is some average viscosity of the
cytoplasm and external solution. In the limit of short wavelengths
($q\rightarrow \infty$) we recover the free bilayer frequency:
$\omega(q)\rightarrow \kappa q^3/4 \eta$.

The mean square amplitude of the normal fluctuations, as a
function of frequency $\omega$, is the Fourier transform
\begin{eqnarray}
d(\omega)^{2}&=&\frac{1}{(2 \pi)^2} \int q dq \int \langle
h_{q}(t)h_{-q}(0)\rangle e^{-i\omega t} dt \nonumber \\
&&=\frac{1}{(2 \pi)^2} \int \langle h_{q}(0)h_{-q}(0)\rangle \frac
{\omega(q)}{\omega(q)^2+\omega^2} q dq
 \label{ampomega}
\end{eqnarray}

For a free bilayer this expression (\ref{ampomega}) gives an
anomalous frequency dependence \cite{anton}: $d(\omega)^{2}\propto
\omega^{-5/3}$. We integrate the expression (\ref{ampomega})
numerically in the range $\pi/R <q<\pi/a$ ($a\simeq 50$\AA), and
compare with the experimental data \cite{atprafi}. In the inset of
Fig.3 we plot $d(\omega)$ using the parameters of the two cells of
Fig.1, and both the high and low values of the effective surface
tension ($\sigma$ and $\sigma_0$ respectively). We find a
reasonable agreement between the calculation and the measurements,
when taking $D_{\omega}\simeq 0.4d$. The similar magnitude of the
bilayer-rigid shell separation from both static and dynamic
experiments, shows the overall consistency of our confinement
model.

In the limit of high frequencies, the earlier result of Brochard
et.al. \cite{brochard}, gave $d(\omega)^{2}\propto \omega^{-4/3}$.
Since $qD_{\omega}\ll 1$ in the measured range, we find in this
limit (\ref{omega}): $\omega(q)\sim \kappa q^5 D_{\omega}^2/4
\eta$, leading to $d(\omega)^{2}\propto \omega^{-7/5}$. It is
difficult to distinguish between these two values using the newer
data \cite{atprafi}. Our calculation has the advantage of
consistently describing both the static (spatial) and dynamic
(temporal) fluctuation data. Note that the case of a pure bilayer
with large effective surface tension $\sigma$, but without the
effect of the rigid wall, is in complete disagreement with the
data (dotted line, Fig.3 inset).

In Fig.3 we show that the normal RBC, ATP depleted RBC and RBC
ghost, are all well described by the same expression
(\ref{ampomega}) (using the softer cell from Fig.1, i.e. the
smaller $\gamma$ and $\sigma$), differing by an effective
temperature factor of up to $\sim3$. This is similar to the
amplitude enhancement factor of $\sim2.5$ found in a previous
study \cite{prost}.

The largest effect of the rigid shell is to increase the effective
viscosity of the water near the bilayer, by constraining its flow.
Defining an effective viscosity by: $\omega=\kappa q^3/4
\eta_{eff}$, we get from (\ref{omega}):
$\frac{\eta_{eff}}{\eta}=\frac{\kappa q^2}{4d^2\left(\gamma+\sigma
q^2\right)}$. At the crossover wavevector $q_{0}$ this function
has its peak: $\eta_{eff}/\eta \simeq 45-30$, depending on the
value of $\sigma$. These values are in close agreement with the
value $\eta_{eff}/\eta \approx 50$ found from the relaxation times
of an electrodeformed RBC \cite{peterson92,sackman84}. In these
experiments, the cytoplasm flows through the cytoskeleton mesh,
setting up a flow field at the crossover wavevector $q_{0}$. Thus,
a rigid cytoskeletal wall, separated by a fixed distance from the
bilayer, accounts for the larger effective viscosity required to
fit these dynamical experiments.

While this model accounts for the wavevector dependence of the
statics and the frequency dependence of the dynamics, the absolute
amplitude of the fluctuations and the different values observed in
active and ATP-depleted cells, must still be explained. One
possibility is that ATP driven fluctuations completely determine
the amplitude of the largest wavelength fluctuations
\cite{atprafi} through the process of spectrin-actin
disconnections and reconnections \cite{atprafi,sackman95,bennett}.
These ATP-driven conformational changes give rise to defects in
the triangular spectrin network, resulting in nodes with more or
less than 6 spectrin bonds. The local curvature of the
cytoskeleton may change at the site of a defect, from being
locally flat (6 bonds) to having a $\sim80$nm deviation out of the
plane of the flat cytoskeleton (5-fold node). The effect of this
random buckling is to increase the mean bilayer-rigid shell
separation by a factor of $\sim 4$. According to our model, this
will increase the amplitude of the $q\rightarrow 0$ modes by a
factor of $\sim 4^4$, as measured \cite{peterson92}.

\begin{acknowledgments}
We thank R. Korenstein, E. Sackmann and H. Strey for useful
discussions. This work was supported by the ISF grant for Center
on Self-Assembly. The authors are grateful to the donors of the
Petroleum Research Fund administered by the American Chemical
Society and to the Schmidt Minerva Center for their support.
N.G.'s research is being supported by the Louis L. and Anita M.
Perlman Postdoctoral Fellowship.
\end{acknowledgments}

\begin{figure}
\centerline{\ \epsfysize 8cm \epsfbox{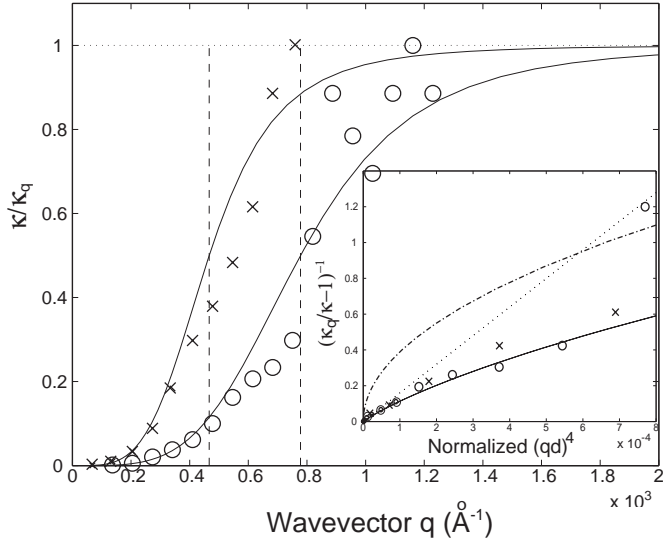}} \caption{The
calculated (Eq. \ref{statcor}) wavevector dependence of the
bending modulus $\kappa/\kappa_q$ of the RBC (solid lines, taking
$\sigma=sigma_0$) compared with the experimental data for the RBC
\cite{sackman87}(o,x). The crossover wavevector $q_{0}$ is
indicated by the vertical dashed lines. Inset: A plot of
$(\kappa_q-\kappa)^{-1}$ as a function of the normalized
wavevector $(qd)^4$ for small wavevectors. The linear slope in the
limit of $q\rightarrow 0$ is indicated by the dotted line. The
deviation from linear behavior is well described by an effective
surface tension $\sigma\simeq\kappa/\xi_{0}^2\sim
2.8,7.7\times10^{-7}$J/m$^2$ for the two cells (solid line). Note
that surface tension alone, without confining wall ($\gamma=0$),
does not describe the data (dash-dot line).}
\end{figure}

\begin{figure}
\centerline{\ \epsfysize 8cm \epsfbox{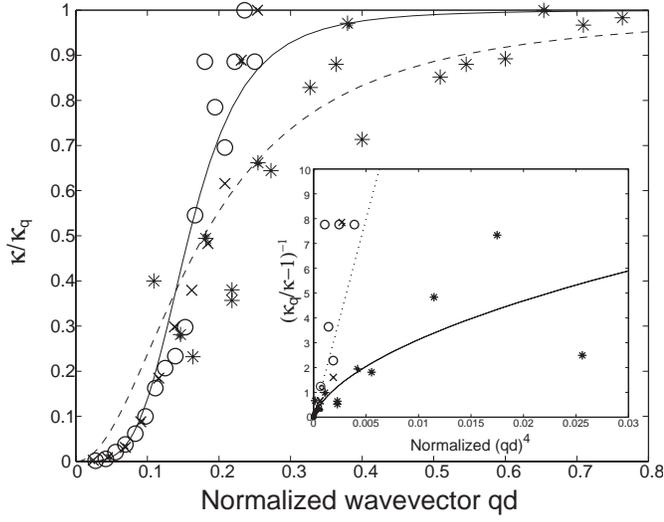}}
 \caption{A plot of the the measured effective modulus
$\kappa/\kappa_q$ of RBC \cite{sackman87}(o,x) and empty giant
vesicles \cite{sackman97}($\ast$) as a function of the normalized
wavevector $qd$. The calculations are given by the solid and
dashed line respectively. Inset: A plot of
$(\kappa_q-\kappa)^{-1}$ as a function of the normalized
wavevector $(qd)^4$ for small wavevectors. The linear slope in the
limit of $q\rightarrow 0$ for the RBC is indicated by the dotted
line. The calculation for the vesicle is given by the solid line.}
\end{figure}

\begin{figure}
\centerline{\ \epsfysize 8cm \epsfbox{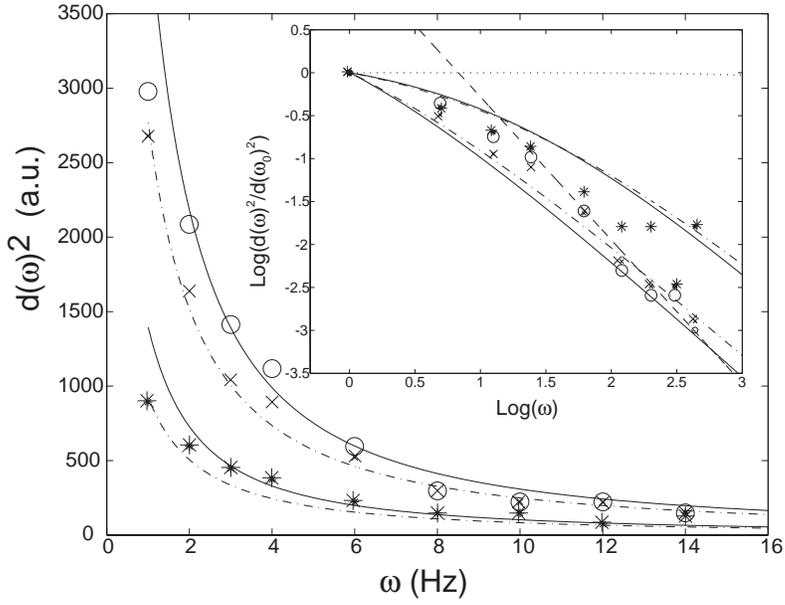}}
% \vskip 5mm
%\centerline{\ \epsfysize 7cm\epsfbox{fig5.eps}}
\caption{Frequency dependence of the mean-square amplitude
\cite{atprafi} $d(\omega)^2$ of the RBC (o),
 showing the reduction in the amplitude due to
partial ATP depletion (x) and complete absence (RBC
ghost)($\ast$). The lines of the calculation (Eq. \ref{ampomega})
differ by the effective temperature enhancement factor of $\sim 3$
(solid lines: $\sigma_0$, dash-dot lines: $\sigma$). Inset: A
normalized Log-Log plot showing the powerlaw dependence
($\omega_{0}=1$Hz). The calculation is done using the parameters
of the two cells of Fig.1 (solid lines: $\sigma_0$, dash-dot
lines: $\sigma$). The dashed line shows the free bilayer behavior
$d(\omega)^2\propto \omega^{-5/3}$. The case of a free bilayer
with large effective surface tension $\sigma$, but without the
effect of the rigid wall, is in complete disagreement with the
data (dotted line).}
\end{figure}

\end{document}